\documentclass[a4paper,11pt]{article}
\usepackage{latexsym}
\usepackage{epsfig, subfigure, subfloat, graphicx, float }
\usepackage{amssymb}
\usepackage{amsmath}
\author{H. Mohseni Sadjadi \footnote{mohsenisad@ut.ac.ir}
\\ {\small Department of Physics, University of Tehran,}
\\ {\small P. O. B. 14395-547, Tehran 14399-55961, Iran}}
\title{Rapid Oscillatory quintessence with variable equation of state parameter in non minimal derivative coupling model}

\begin{document}
\maketitle
\begin{abstract}
The rapid oscillating scalar field is considered as the
quintessence in the framework of nonminimal kinetic coupling
model. The scalar field behaves like a perfect fluid with a
variable equation of state parameter which can be expressed
as a function of the Hubble parameter. This feature enables us to study its evolution via continuity equation.
 The scalar field may behave as dark energy in the present epoch, while
it behaves like dark matter in the future. This characteristic allows the
occurrence of a decelerated expansion after the present acceleration, without invoking any interaction.
The model is also studied in the presence of an interaction between dark sectors
alleviating the coincidence problem. The stability of the model is investigated and
the stable attractor solutions are studied.
\end{abstract}

\section{Introduction}
In 1998,  based on astrophysical data,  some evidences
indicating that the expansion of the Universe has been accelerated
\cite{acc}, were found. Many models have been proposed to describe this odd
phenomenon which is still inexplicable in the context of the
general reactivity with known matters.  These models may be classified in two categories:
(Einstein) General theory of gravity with an exotic source with negative pressure dubbed as dark energy \cite{dark energy},
and theories based on modification of general relativity in large scale \cite{modified}.
The most natural and simple candidate for dark energy
is the cosmological constant which
suffers from fine tuning and coincidence problems,
and may not be related to the vacuum energy of the standard particle physics model due to its
extremely small value \cite{CC}.
Inspired by the inflation theory in the early Universe, one can introduce an
 exotic scalar field dubbed as quintessence
playing the role of dark energy \cite{quint}. The advantage of this model is
its dynamics which may solve the coincidence problem
and explain the evolution of dark energy density.
In this framework, to describe the present acceleration
of the Universe, the scalar field model with a non-minimal derivative
coupling to gravity was proposed in \cite{sush}. Thereafter, the non minimal derivative coupling model has also
been employed to describe the possible super acceleration in
the late time \cite{sad}. Further studies about this model can be
found in \cite{nonminimal}.

Although most of the scalar field models use the slow roll
conditions to explain the inflation, but as was quoted in
\cite{rapidinfl}, the acceleration may be continued in the rapid
oscillation after the slow roll.  Also in \cite{rapidquin} it was
tried to introduce the quintessence as a rapid oscillating scalar
field. In the rapid oscillation phase the scalar field behaves as a
fluid with a {\it {constant equation of state (EoS) parameter}} which, by a
suitable choice of the potential, becomes enough negative to result
in the cosmological acceleration.

In this paper we consider the spatially flat Friedman Robertson
Walker (FRW) Universe in the framework of non minimal derivative
coupling model proposed in \cite{sush}.

In the second section we give an overview of the model and
introduce the rapid oscillating scalar field as the quintessence.
As in the present time we cannot neglect the contribution of dark matter, we generalize the results obtained in \cite{sad1}
which was derived for a Universe consisted only of
inflaton. In contrast to the minimal case,
the scalar field acquires a dynamical  EoS parameter which depends on the Hubble parameter. This enable us to use the continuity equation to study 
its evolution. The dependence of the EoS parameter on the Hubble parameter 
gives rise to some novel and interesting features such as the
possibility that the Universe reenters a decelerated expansion without considering
any energy exchange between dark matter and dark energy components. This feature is explained explicitly
by deriving a series solution in the third section.  In that section we also obtain the ratio of dark energy and dark matter energy densities at the beginning
of the acceleration (deceleration) phase. In resemblance to the inflaton decay
during its rapid oscillation, we consider also the case that the
oscillatory quintessence is allowed to exchange energy but only
with dark matter alleviating the coincidence problem. The stability of the model and its late time attractor solutions are discussed. At the end of the third  section a short discussion is given on the instabilities of interacting dark energy model in the perturbed FRW Universe.
In the fourth section, our previous results are discussed by using some illustrative examples. The evolution of dark energy and dark energy are numerically depicted in terms of the redshift parameter and behaviors of the EoS parameter of the scalar field and the decelerated parameter of the Universe are discussed.

We use units $\hbar=c=1$.

\section{Rapid oscillatory scalar field with variable EoS parameter}
We consider the action
\begin{equation}\label{1}
S=\int\left[{1\over 2}M_{P}^2R-{1\over 2}(g^{\mu \nu}-WG^{\mu
\nu})\partial_{\mu}\varphi
\partial_{\nu}\varphi-V(\varphi)\right]\sqrt{-g}d^4x+S_m,
\end{equation}
describing a scalar field, $\varphi$, whose the kinetic term is
coupled to the Einstein tensor $G^{\mu \nu}=R^{\mu \nu}-{1\over
2}g^{\mu \nu}R$ via the coupling  $W$ which has inverse-square
mass dimension \cite{sush}. $S_m$ is the matter action containing components
other than the scalar field. $M_P$ is the reduced Planck mass,
$M_P=\sqrt{1\over 8\pi G}\approx 1.2209\times 10^{19}GeV$. The
Universe is taken to be a spatially flat FRW space time, with
scale factor $a(t)$.

In the absence of interaction between $\varphi$ and other sectors, the equation of
motion of the scalar field $\varphi(t)$ is
\begin{equation}\label{2}
(1+3WH^2)\ddot{\varphi}+3H(1+3WH^2+2W\dot{H})\dot{\varphi}+V'(\varphi)=0,
\end{equation}
where $H:={\dot{a}\over a}$ is the Hubble parameter. The dark
energy density, $\rho_{\varphi}$,  and pressure, $P_{\varphi}$,
are given by
\begin{eqnarray}\label{3}
\rho_{\varphi}&=&{1\over
2}(1+9WH^2)\dot{\varphi}^2+V(\varphi)\nonumber \\
P_{\varphi}&=&{1\over 2}(1-3WH^2)\dot{\varphi}^2-V(\varphi)-W{d
(H\dot{\varphi}^2)\over dt},
\end{eqnarray}
satisfying the continuity equation
\begin{equation}\label{r1}
\dot{\rho_\varphi}+3H(\rho_\varphi+P_\varphi)=0.
\end{equation}
In our study, $V(\varphi)$ is an even function specified by the
power law form
\begin{equation}\label{4}
V(\varphi)=\lambda \varphi^n,
\end{equation}
where $n>0$. We focus on (quasiperiodic) rapidly oscillating scalar field solutions
characterized by
\begin{equation}\label{5}
H\ll {1\over T},
\end{equation}
where $T$ is the period of the $\varphi$ oscillation, and
${\dot{H}\over H}\ll {1\over T}$ . $T$ is given by
\begin{equation}\label{r3}
T=2\int_{-\Phi}^{\Phi}{d\varphi\over \dot{\varphi}},
\end{equation}
where $\Phi$ is the amplitude of quasiperiodic oscillation of
$\varphi$.   This type of solution was
discussed in the literature to study the reheating era after the
inflation in the minimal coupling case \cite{rapid} .
Using the Friedman equation
\begin{equation}\label{r2}
H^2={1\over 3M_P^2}\rho,
\end{equation}
we find that the total energy
density, like the Hubble parameter, does not change significantly
during one oscillation \cite{rapid}
\begin{equation}\label{r4}
\left|2{\dot{H}\over H}\right|=\left|{\dot{\rho}\over
\rho} \right|\ll {1\over T(t)}.
\end{equation}

In the non minimal derivative coupling case (\ref{1}) and
in high friction regime: $WH^2\gg 1$, the
rapidly oscillating scalar field, following
the same steps as \cite{rapid}, was studied in details in \cite{sad1}(we do not
repeat the details here) in a Universe dominated by $\varphi$ and in high friction regime $WH^2\gg 1$, the solution may be represented as
$\varphi(t)=\Phi(t)\cos\left({2\pi t\over T(t)}\right)$,
where $\Phi$ is the amplitude of the $\varphi$ oscillation.
Like the harmonic oscillation the energy may be approximated by
the energy potential at the tuning points:
\begin{equation}\label{r5}
\rho_\varphi=V(\Phi)=\lambda \Phi^n.
\end{equation}

The period of an oscillation is given (see \cite{rapid})
\begin{eqnarray}\label{6}
T&=&2\int_{-\Phi}^{\Phi}{d\varphi \over \dot{\varphi}}\nonumber \\
&=&\sqrt{2}\int_{-\Phi}^{\Phi}{\sqrt{1+9WH^2} \over
\sqrt{\rho_{\varphi}-V(\varphi)}}d\varphi\nonumber \\
&=&2\sqrt{1+9WH^2\over \lambda}\Phi^{1-{n\over 2}}\int_{0}^{1}{dx\over \sqrt{1-x^n}}\nonumber \\
&=&{\Gamma\left({1\over n}\right)\over n\Gamma\left({1\over
2}+{1\over n}\right)}\sqrt{8\pi(1+9WH^2)\over
\lambda}\Phi^{1-{n\over 2}}.
\end{eqnarray}
Rapid oscillation condition (\ref{5}) may now be rewritten as
\begin{equation}\label{7}
{\Phi \over M_P}\ll \sqrt{3\over 8\pi}{n\Gamma\left({1\over 2}+{1\over
n}\right)\over \Gamma\left({1\over n} \right)}\sqrt{\Omega_{\varphi}\over {1+9WH^2}},
\end{equation}
where $\Omega_\varphi={\rho_\varphi\over 3M_P^2 H^2}$. Note that
$\Omega_\varphi$ cannot exceed unity. For $WH^2\ll 1$
(\ref{7}) reduces to the result obtained in minimal case
\cite{rapidquin}, and for high friction regime $WH^2\gg 1$  we
obtain
\begin{equation}\label{8}
\Phi\ll \left(\sqrt{1\over 8\pi}{n\Gamma\left({1\over 2}+{1\over
n}\right)\over \Gamma\left({1\over n} \right)}\right)^{2\over
n+2}\left({M_P^2\Omega_{\varphi}\over
\sqrt{W\lambda}}\right)^{2\over n+2},
\end{equation}
which is the same as what was obtained in \cite{sad1} for $\Omega_\varphi= 1$.
In general (\ref{5}) is satisfied when
\begin{equation}\label{9}
\lambda\gg  \left({\Gamma\left({1\over n} \right)\over
n\Gamma\left({1\over 2}+{1\over n}
\right)}\right)^n\left(8\pi(9\tilde{W}\tilde{H}^2+1)\right)^{n\over
2}\left({H_0\over
M_P}\right)^2(3\Omega_\varphi)^{2-n\over2}\tilde{H}^2 M_P^{4-n},
\end{equation}
where the dimensionless parameters are defined by
$\tilde{W}=WH_0^2$ and $\tilde{H}={H\over H_0}$. $"0"$ indicates
the present time. The (time) average of a quantity over an
oscillation whose period is $T$ is given by
\begin{equation}\label{10}
\left<A(t)\right>={\int_t^{t+T}A(t')dt'\over T}.
\end{equation}

\begin{figure}[h]
\centering\epsfig{file=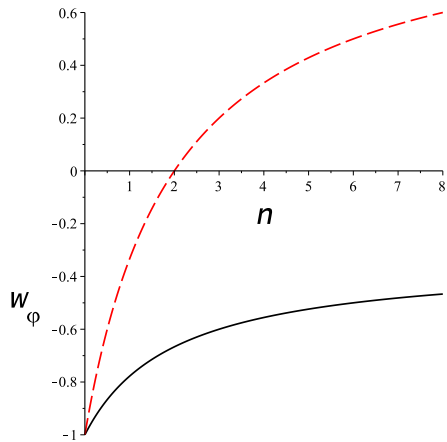,width=6cm,angle=0} \caption{ Equation
of state parameter of the rapid oscillating scalar field depicted
in terms of $n$ for minimal coupling (red dashed line) and  nonminimal coupling in high friction regime (black solid
line). In the nonminimal case, for all positive $n$, we have $w_\varphi\leq -{1\over 3}$. } \label{fig0}
\end{figure}

The adiabatic index of the scalar field, $\gamma_\varphi$ is defined by
\begin{equation}\label{13}
\gamma_\varphi:=1+{<P_\varphi>\over <\rho_\varphi>}.
\end{equation}
Using (\ref{2}), we obtain \cite{sad1}
\begin{eqnarray}\label{14}
\gamma_\varphi&=&{\left<(1+3WH^2)\dot{\varphi}^2-W{d\over
dt}(H\dot{\varphi}^2)\right>\over <\rho_\varphi>}\nonumber \\
&=&{2(1+3WH^2)\over
(1+9WH^2)V(\Phi)}{\int_{-\Phi}^{\Phi}\sqrt{\rho_\varphi
-V(\varphi)}d\varphi\over \int_{-\Phi}^{\Phi}{d\varphi\over
\sqrt{\rho_\varphi-V(\varphi)}}}\nonumber \\
&=&{2(1+3WH^2)\over 1+9WH^2}{n\over n+2}.
\end{eqnarray}
The EoS parameter of the scalar field field is then
\begin{equation}\label{16}
w_\varphi=\gamma_\varphi-1={n-2-3(n+6)WH^2\over (n+2)(1+9WH^2)}.
\end{equation}
In the minimal case, $WH^2\ll 1$, we recover the very well known
relation \cite{kolb}
\begin{equation}\label{17}
w_\varphi={n-2\over n+2}.
\end{equation}
The scalar field mimics the behavior of barotropic fluid with a
constant EoS parameter. For $WH^2\gg 1$,  we recover our previous result \cite{sad1}
\begin{equation}\label{18}
w_\varphi=-{n+6\over 3n+6}.
\end{equation}
For a specific $n>0$, EoS parameter in high friction regime,
$WH^2\gg 1$, is smaller than that in $WH^2\ll 1$ limit (see fig.(\ref{fig0})).
\section{Evolution equations}

By taking the time average of the continuity equation
(\ref{r1})
over an oscillation, and bearing in mind that $H$ changes very
slowly in one period, we obtain
\begin{equation}\label{r6}
<\dot{\rho_\varphi}>+3H <\rho_\varphi>\left(1+{<P_\varphi>\over
<\rho_\varphi>}\right)=0.
\end{equation}
This can be regraded as the continuity equation for the effective energy density $<\rho_\varphi>$
in time scales
larger than the period of oscillation, where the replacement of
parameters by their average values is acceptable \cite{kolb}.\footnote{
Indeed within a period, some of the parameters such as $P_\varphi$
are different from their average values.} By definition we have
\begin{eqnarray}\label{11}
\left<\dot{\rho_\varphi}(t)\right>&=&{\rho_\varphi(t+T)-\rho_{\varphi}(t)\over
T}\nonumber \\
&=&\dot{\rho_\varphi}(t),
\end{eqnarray}
where the last equality is valid for $t\gg T$. Finally the continuity equation becomes
\begin{equation}\label{15}
\dot{\rho_\varphi}+6H{1+3WH^2\over 1+9WH^2}{n\over
n+2}\rho_\varphi=0.
\end{equation}
Hence the scalar field behaves as a perfect fluid with known equation of state parameter
in terms of the Hubble parameter.
Henceforth we assume that the Universe is nearly filled with a
rapid oscillating scalar field dark energy and matter which is
composed of baryonic, $\rho_b$, and also cold dark matter
$\rho_{cd}$
\begin{equation}\label{21}
\rho_m=\rho_{cd}+\rho_b\simeq \rho_{cd}.
\end{equation}
The Friedman equation is then
\begin{equation}\label{22}
H^2={1\over 3M_P^2}(\rho_m+\rho_\varphi).
\end{equation}
In the absence of  interaction between the dark energy and other sectors we can write
\begin{eqnarray}\label{rr10}
&&\dot{\rho_\varphi}+{1+3WH^2\over 1+9WH^2}{6nH\over
n+2}\rho_\varphi=0 \nonumber \\
&&\dot{\rho_m}+3H\rho_m=0.
\end{eqnarray}
The Hubble parameter is a decreasing function of time and its infimum  is $0$. Therefore we expect that $H$ asymptotes to $0$ or a constant at late time
$\lim H_{t\to \infty }=C$, where $C$ is a nonnegative constant. One can show that $C=0$: Indeed putting $H=C$  in (\ref{rr10}) reveals that both $\rho_\varphi$ and $\rho_m$ tends to $0$ at late time. But From Friedman equation this leads to $H\to 0$ .

\subsection{Accelerated expansion}
Obtaining an analytical solution for (\ref{rr10}) is not feasible. But one can examine the ability of the model to describe  the deceleration to acceleration phase transition and its converse (which is salient feature of this model), by presenting a series solution near the possible transition time. When the sign of $\dot{H}+H^2$ changes from negative to positive,
the decelerated expansion of the Universe becomes accelerated. By
assuming that the Hubble parameter is an analytic function of
time,  in the neighborhood of the transition time, which for
simplicity we take as $t=0$, the Hubble parameter has the Taylor
expansion $H=h_0+h_1
t^\alpha+h_2t^{\alpha+1}+\mathcal{O}(t^{\alpha+1})$. $\alpha$ is the order of the first
nonzero derivative of $H$ with respect to $t$ at transition time.
 $\dot{H}+H^2=0$ at $t=0$, leads us to take $\alpha=1$ and $h_1=-h_0^2$.
Hence near the transition time we have
\begin{equation}\label{rr12}
H=h_0-h_0^2t+h_2t^2+\mathcal{O}(t^3).
\end{equation}
In this way $\dot{H}+H^2=2(h_2-h_0^3)t$. For $h_2>h_0^3$ the
transition occurs from deceleration to acceleration phase at $t=0$. The
opposite transition occurs for  $h_2<h_0^3$. To see whether our
model admits such a solution, we examine  (\ref{rr12}) in the continuity and
Friedman equations. With the Hubble parameter taken as (\ref{rr12}), the
series solution of the $\rho_m$ is given by
\begin{equation}\label{rr13}
\rho_m(t)=\rho_m(0)-3h_0\rho_m(0)t+6h_0^2\rho_m(0)t^2+\mathcal{O}(t^3).
\end{equation}
In the same way
\begin{eqnarray}\label{rr14}
&&\rho_\varphi(t)=\rho_\varphi(0)-{6n(1+3Wh_0^2)h_0\rho_\varphi(0)\over
(n+2)(1+9Wh_0^2)}t\nonumber
\\&&+{3nh_0^2(7n+36nWh_0^2+81nW^2h_0^4+54W^2h_0^4+2)\rho_\varphi(0)\over
(1+9Wh_0^2)^2(n+2)^2}t^2\nonumber \\
&&+\mathcal{O}(t^3).
\end{eqnarray}
By inserting (\ref{rr12}), (\ref{rr13}), and (\ref{rr14}) in the Friedman equation and equating
the coefficients of each power of $t$, we deduce that this
equation, up to $\mathcal{O}(t^3)$, is satisfied provided that:
\begin{equation}\label{rr15}
h_0^2={1\over 3M_P^2}(\rho_\varphi(0)+\rho_m(0)),
\end{equation}
\begin{equation}\label{rr16}
h_0^2={1\over 2M_P^2}\left(\rho_m(0)+{2n(1+3Wh_0^2)\over
(n+2)(1+9Wh_0^2)}\rho_\varphi(0)\right),
\end{equation}
and
\begin{eqnarray}\label{rr17}
&&2h_2={1\over
3M_P^2}\Big({3nh_0(7n+36nWh_0^2+81nW^2h_0^4+54W^2h_0^4+2)\over
(n+2)^2(1+9Wh_0^2)^2}\rho_\varphi(0)\nonumber \\
&&+6h_0\rho_m(0)\Big)-h_0^3
\end{eqnarray}
hold. Using these equations we obtain
\begin{equation}\label{rr18}
h_2-h_0^3=
{-h_0^3\left(-n^2+3n-2+9(n^2+4n-4)Wh_0^2-27(n+6)W^2h_0^4\right)\over
(n+2)(1+9Wh_0^2)\left(-n+2+3(n+6)Wh_0^2\right)}
\end{equation}
In the minimal model $W=0$, $h_2-h_0^3>(<)0$ reduces to
$-{n-1\over n+2}>(<)0$. So {\it{in this case there may be only one
transition}}. In the nonminimal case, the situation changes and
$h_2-h_0^3>(<)0$ is equivalent to
\begin{equation}\label{s1}
{-n^2+3n-2+9(n^2+4n-4)Wh_0^2-27(n+6)W^2h_0^4\over
(n+2)\left(-n+2+3(n+6)Wh_0^2\right)}<(>)0.
\end{equation}
Therefore, in principle,  we may have
both deceleration to acceleration and acceleration to deceleration
phase transitions.

The ratio of energy densities at transition time is given by
\begin{equation}\label{rr19}
r={\rho_\varphi(0)\over \rho_m(0)}={n+2+9nWh_0^2+18Wh_0^2\over
4(9Wh_0^2-n+1)}
\end{equation}
For $Wh_0^2\gg 1$, $r\simeq {1\over 2}+{n\over 4}$ while for
$Wh_0^2\ll 1$ we have $r={n+2\over 4(1-n)}$.

\subsection{Interacting case and attractor solutions}

We can generalized  the model by allowing energy exchange between dark components.
In the presence of an interaction $Q$, the continuity equations
read
\begin{eqnarray}\label{23}
&&\dot{\rho_\varphi}+{1+3WH^2\over 1+9WH^2}{6nH\over
n+2}\rho_\varphi=Q \nonumber \\
&&\dot{\rho_m}+3H\rho_m=-Q.
\end{eqnarray}
Inspired by the models proposed for reheating in which the
inflation decays during coherent rapid oscillation \cite{kolb}, we
choose the source term as $Q=-\Gamma
\dot{\varphi}^2,\,\,\Gamma>0$, but here instead of radiation
(ultra relativistic particles) the scalar field is assumed to
decay to dark matter. So the equation of motion of the scalar
field becomes
\begin{equation}\label{24}
(1+3WH^2)\ddot{\varphi}+3H(1+3WH^2+2W\dot{H})\dot{\varphi}+V'(\varphi)=-\Gamma
\dot{\varphi},
\end{equation}
which in the minimal case reduces to the well known equation for
inflaton decay \cite{kolb}
\begin{equation}\label{25}
\ddot{\varphi}+3H\dot{\varphi}+V'(\varphi)=-\Gamma \dot{\varphi}.
\end{equation}
With the same method used to compute $\gamma_\varphi$ in
(\ref{14}), we arrive at
\begin{equation}\label{26}
<Q>=-\Gamma<\dot{\varphi}^2>=-{2n\Gamma\over
(n+2)(1+9WH^2)}\rho_\varphi.
\end{equation}
An interesting property of this interaction is that it depends on
$H$ such that it is inoperative for $\Gamma \ll (1+9WH^2)$. But,
as $\dot{H}<0$, $H$ decreases during Universe expansion and the
decay becomes more relevant. This may alleviate the coincidence
problem, i.e. why the matter and dark energy densities are at the
same order today. This energy exchange, postpones
the beginning of the acceleration phase of the Universe and also expedites
the possible dark matter domination at late time.
Let us now study  the late time attractor solutions of the model and investigate the fate
of the Universe.

By defining dimensionless parameters $\tau:=tH_0$, $\tilde{H}:={H\over H_0}$, $\tilde {W}:=WH_0^2$, $\tilde{\Gamma}:={\Gamma\over H_0}$, and also dimensionless fractional energy densities $\Omega_m:={\rho_m\over 3M_P^2H^2}$, $\Omega_\varphi:={\rho_\varphi\over 3M_P^2 H^2}$, the continuity equations and the Friedman equation after some computations reduce to a system of autonomous differential equations
\begin{eqnarray}\label{w1}
&&\Omega_\varphi'=-{2n\tilde{\Gamma}\over (n+2)(1+9\tilde{W}\tilde{H}^2)}\Omega_\varphi-\left({6n\tilde{H}\over n+2}\right)\left({1+3\tilde{W}\tilde{H}^2\over 1+9\tilde{W}\tilde{H}^2}\right)\Omega_\varphi\nonumber \\
&&+3\tilde{H}\left(\Omega_m+\left({2n\over n+2}\right)\left({1+3\tilde{W}\tilde{H}^2\over 1+9\tilde{W}\tilde{H}^2}\right)\Omega_\varphi\right)\Omega_\varphi\nonumber \\
&&\Omega_m'=-3\tilde{H}\Omega_m+3\tilde{H}\left(\Omega_m+\left({2n\over n+2}\right)\left({1+3\tilde{W}\tilde{H}^2\over 1+9\tilde{W}\tilde{H}^2}\right)\Omega_\varphi\right)\Omega_m\nonumber \\
&&+{2n\tilde{\Gamma}\over (n+2)(1+9\tilde{W}\tilde{H}^2)}\Omega_\varphi\nonumber \\
&&\tilde{H}'=-{3\over 2}\tilde{H}^2\left(\Omega_m+\left({2n\over n+2}\right)\left({1+3\tilde{W}\tilde{H}^2\over 1+9\tilde{W}\tilde{H}^2}\right)\Omega_\varphi\right)
\end{eqnarray}
Prime denotes derivative with respect to the time $\tau$. Note that as the EoS parameter of the scalar field and the interaction term are $H$ dependent, we cannot simply eliminate $H$ in our equations by redefinition of time parameter, hence the evolution of the Hubble parameter is included in the three dimensional autonomous system.

By setting $\Omega_\varphi'=0,\,\,\, \Omega_m'=0,\,\,\,\tilde{H}'=0$ we can find the fixed points $(\Omega_m^*, \Omega_\varphi^*, H^*)$. The only fixed point consistent with $\Gamma\neq 0$ is  $(\Omega_m^*=1, \Omega_\varphi^*=0, H^*=0)$.  To study the stability of this fixed point we consider small homogenous variations about the fixed point:
$ \Omega_\varphi^*+\delta \Omega_\varphi$,  $ \Omega_m^*+\delta \Omega_m$,  $H^*+\delta H$. Inserting these into (\ref{w1}) we obtain
\begin{equation}\label{w2}
{d\over d\tau}\left(
\begin{array}{c}
\delta \Omega_\varphi \\
\delta \Omega_m \\
\delta H \\
\end{array}
\right)=\mathcal{M} \left(
\begin{array}{c}
\delta \Omega_\varphi \\
\delta \Omega_m \\
\delta H \\
\end{array}
\right)
\end{equation}
where $\mathcal{M}=\left( \begin{array}{ccc} -{2n\over n+2} \tilde{\Gamma}& 0 & 0 \\ {2n\over n+2} \tilde{\Gamma} & 0 & 0 \\
0 & 0 & 0 \\ \end{array} \right)$. The Jordan form of $\mathcal{M}$ is given by\newline
 $\left( \begin{array}{ccc} 0& 0 & 0 \\ 0 & -{2n\over n+2} \tilde{\Gamma} & 0 \\
0 & 0 & 0 \\ \end{array} \right)$, therefore as $n>0$ by definition, the critical point is unstable only for $\Gamma<0$.
So in our model there is a late time stable attractor associated to cold matter dominated epoch. In this framework, the present acceleration
of the Universe must be considered as a transient era.
\subsection{Interacting dark energy in perturbed FRW Universe}
Cosmological perturbations in the inflationary era in a model where the rapid oscillating scalar field
is considered as the inflaton (the only ingredient of the early Universe), is discussed in details in \cite{sad1}.
Here we are only concerned about the late time acceleration where other ingredients are also present.
Requiring stability against perturbations puts some constraints on
the parameters. In this part based on \cite{Maartens0,Maartens1} we study briefly this subject for some special cases.

Consider the perturbed FRW space time
\begin{equation}\label{rr1}
ds^2=a^2\left(-(1+2\xi)d\eta^2+2\partial_iB d\eta
dx^i+((1-2\psi)\delta_{ij}+2\partial_i\partial_jE)dx^idx^j\right).
\end{equation}
$\eta$ is the conformal time and $\xi$, $B$, $\psi$, and $E$ are
scalar functions. The $nth$ fluid four vector velocity is
$u_n^\nu=a^{-1}(1-\xi,\overrightarrow{\partial} v_n)$ where $v_n$
is the peculiar velocity potential. Evolution of the density
perturbation of the $nth$ component, $\delta_n:={\delta
\rho_n\over \bar{\rho}_n}$, and velocity perturbation
$\theta_n=-k^2(B+v_n)$ are given by \cite{Maartens0}
\begin{eqnarray}\label{rr2}
&&\delta_n'+3\mathcal{H}(c_{sn}^2-w_n)\delta_n+\gamma_n\theta_n+3\mathcal{H}\left( 3\mathcal{H}\gamma_n(c_{sn}^2-w_n)+w_n'\right){\theta_n\over k^2}\nonumber \\
&&+k^2\gamma_n(B-E')-3\gamma_n\psi'={a\bar{Q}_n\over
\bar{\rho}_n}\left(\xi-\delta_n+3\mathcal{H}(c_{sn}^2-w_n){\theta_n\over
k^2} \right)\nonumber \\
&&+{a\over \bar{\rho}_n}\delta Q_n,
\end{eqnarray}
\begin{eqnarray}\label{rr3}
&&\theta_n'+\mathcal{H}(1-3c_{sn}^2)\theta_n-{k^2c_{sn}\over \gamma_n}\delta_n+{2k^4\pi_n\over 3a^2\gamma_n\rho_n}-k^2\xi\nonumber \\
&&=
{a\over \bar{\gamma}_n \bar{\rho}_n}(\bar{Q}_n
(\theta-(1+c_{sn}^2)\theta_n)
- k^2f_n)
\end{eqnarray}

Prime denotes derivative with respect to the conformal time and
$\mathcal{H}={d lna\over d\eta}$. Bar denotes the unperturbed value of a parameter on the background, e.g. $\rho_n=\bar{\rho}_n+\delta \rho_n$. $\pi_n$ is related to
anisotropic stress tensor $\pi_{n
i}^j=(\partial^j\partial_i-{1\over
3}\delta^{i}_{j}\nabla^2)\pi_n$. $c_{sn}={\delta P_n\over \delta
\rho_n}$ is the sound speed in the rest frame of $nth$ fluid.
$f_n$ is the momentum transfer potential. Each fluid has its own
evolution equation $\rho_n'+3\mathcal{H}\gamma_n\rho_n-aQ_n=0$. For
a system of interacting dark energy $\rho_\phi$  and dark matter
$\rho_m$, $Q_m=-Q_\varphi=-Q$. $\theta=-k^2(v+B)$, where $v$ is the
total four velocity potential.

For $w_n\rightarrow -1$, the equation (\ref{rr3}) exhibits
an instability. For the interaction $Q=\Gamma \dot{\varphi}^2$, we have ${Q\over \gamma_\varphi}={\Gamma\over 1+3WH^2}$. In large scale  $\xi$ and $\delta$ terms may be dropped and for $(3WH^2+1)\gg \Gamma$ the interaction term may be neglected and there is no instability \cite{Maartens1}.  For $WH^2\ll 1$, solving (\ref{rr3}),  one obtains \cite{Maartens1}
\begin{equation}\label{rr4}
\theta_\varphi=\theta_\varphi(\Gamma=0)e^{\beta  \Gamma(t-t_0)}.
\end{equation}
$\beta=2 (1)$, when the transfer of the energy
momentum is parallel to  dark matter (dark energy) and $t_0$ is the time at which  $\theta_\varphi=\theta_\varphi(\Gamma=0)$.  $\Gamma>0$  leads to an instability which has significant fingerprints in the present era provided that the growth rate of velocity perturbation satisfies  $\beta \Gamma \gtrsim H_0$.

Note that the above discussions are true in the domain of validity of  $WH^2\gg 1$ and $WH^2\ll 1$.
Other interactions may also be taken  for this theory, such as $Q\propto \rho_m$ which was considered in \cite{Maartens2}. The large scale non adiabatic instability in this case may be cured by allowing dark energy EoS parameter to vary \cite{Maartens2}. In our model, this variation is not an ad hoc assumption but a characteristic of our solution.

\section{Numerical illustrations and discussions }

The continuity equations of energy densities, in terms of the redshift parameter $z$ defined by $a={1\over 1+z}$, are
\begin{eqnarray}\label{27}
&&-(1+z){d\over
dz}\tilde{\rho_\varphi}+3\gamma_\varphi\tilde{\rho_\varphi}=-\left({2n\over
n+2}\right){\tilde{\Gamma}\over
\tilde{H}(1+9\tilde{W}\tilde{H}^2)}\tilde{\rho_\varphi}\nonumber \\
&&-(1+z){d\over dz}\tilde{\rho_m}+3\tilde{\rho_m}=\left({2n\over
n+2}\right){\tilde{\Gamma}\over
\tilde{H}(1+9\tilde{W}\tilde{H}^2)}\tilde{\rho_\varphi}\nonumber \\
&&-(1+z){d\over dz}\tilde{\rho_b}+3\tilde{\rho_b}=0.
\end{eqnarray}
$\gamma_\varphi$ is given by eq.(\ref{14}). The energy
exchange is considered only between dark sectors.
the Hubble parameter is given by $\tilde{H}^2=\tilde{\rho_\varphi}+\tilde{\rho_{cd}}+\tilde{\rho_b}$. We have employed
dimensionless parameters specified by a tilde character:
$\tilde{\Gamma}={\Gamma\over H_0} \,\,\,$,
 $\tilde{W}=WH_0^2\,\,\,$,
 $\tilde{H}={H\over H_0}\,\,\,$, $\tilde{\rho_i}={\rho_i\over
\rho_c(0)}$ where the present critical energy density is
$\rho_c(0)=3M_P^2 H_0^2\,\,\,$, and"0" as before denotes the
present time characterized by  $a=1$ or $z=0$. Therefore the
initial conditions are $\tilde{H}(0)=1$ and
$\tilde{\rho_i}(0)=\Omega_i(0)$, where $\Omega_i={\rho_i\over
3M_P^2 H^2}$.

In the absence of an analytical solution for the system (\ref{27}), let us use numerical
methods to illustrate the evolution of the model and justify our previous results. In our analysis we
neglect the contribution of the radiation.

\begin{figure}[H]
\centering\epsfig{file=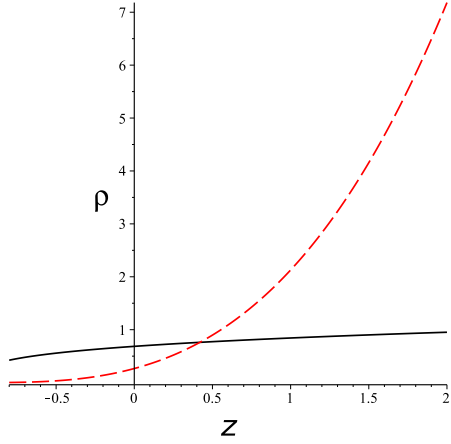,width=6cm,angle=0}
\caption{$\tilde{\rho_m}$ (red dashed line) and
$\tilde{\rho_{\varphi}}$ (black solid line) depicted in terms of the redshift $z$ for
the parameters
$\{\tilde{W}=1000,n=0.352,\Gamma=0,\Omega_{\varphi}(0)=0.685,
\Omega_{cd}(0)=0.266, \Omega_b(0)=0.049\}$. $\varphi$ has a nearly constant EoS parameter, therefore $\tilde{\rho}_{\varphi}$ decreases very slowly.} \label{fig1}
\end{figure}

In fig.(\ref{fig1}), dark energy and cold dark matter densities
are depicted in terms of the redshift parameter for
$\{\tilde{W}=1000,w_\varphi(0)=-0.9,\Omega_{\varphi}(0)=0.685,
\Omega_{cd}(0)=0.266, \Omega_b(0)=0.049\}$, in the absence of
interaction. The  values of fractional energy densities and the nearly constant EoS parameter of dark energy
at the present time are picked from Planck 2013 data \cite{Planck}. In this case the potential is $V(\varphi)=\lambda
\left|\varphi\right|^n$ \cite{rapidinfl}. For $z> 0.5$ the matter was dominated but as the EoS
parameter of the scalar field is negative, dark energy becomes
dominant later.
The EoS parameter of $\varphi$ is approximately a constant: $w_\varphi\simeq -1$, therefore
as it can be seen in fig.(\ref{fig1}), in the absence of interaction $\rho_\varphi$
changes very slowly.

\begin{figure}[H]
\centering\epsfig{file=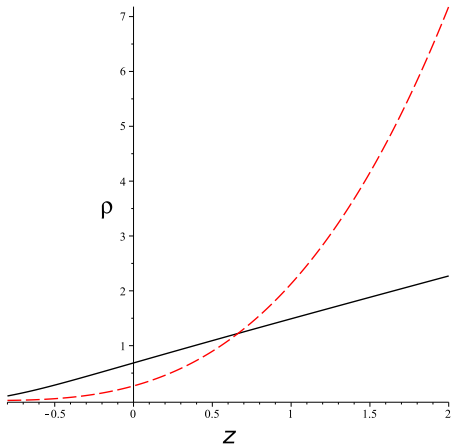,width=6cm,angle=0} \caption{
$\tilde{\rho_{cd}}$ (red dashed line) and $\tilde{\rho_{\varphi}}$
(solid black line) in terms of $z$ and for $\{\tilde{W}=1, n=2,
\Gamma=0, \Omega_{\varphi}(0)=0.685, \Omega_{cd}(0)=0.266,
\Omega_b(0)=0.049 \}.$ $\varphi$ has not a constant EoS parameter, and ultimately behaves as pressureless matter. $\rho_\varphi$ is diluted faster than
the high friction case in the previous figure.} \label{fig2}
\end{figure}

In the intermediate regime where $WH^2\gg 1$ does not hold,  $\varphi$ has a dynamical EoS parameter. In this
case the behaviors of $\tilde{\rho}_\varphi$ and $\tilde{\rho}_{cd}$, are plotted
in fig.(\ref{fig2}), for $\tilde{W}=1$ and a quadratic potential.

\begin{figure}[H]
\centering\epsfig{file=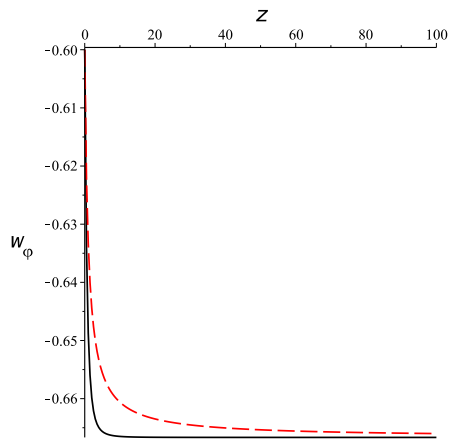,width=6cm,angle=0} \caption{EoS
parameter of oscillatory dark energy in our model (black solid line) with the same initial conditions as in fig.(\ref{fig2}), and  Chevallier-Polarski-Linder parametrization $w_\varphi=w_\varphi(0)+w_{\varphi a} (1-a)$ (red dashed line), with
$\{\omega[0]=-0.6,\omega[a]=-0.066\}$ depicted in terms of $z$.}
\label{fig3}
\end{figure}

In this example, the EoS parameter shows a good agreement with the Chevallier-Polarski-Linder parametrization \cite{CPL},
\begin{equation}\label{cpl}
w_\varphi=w_\varphi(0)+w_{\varphi a} (1-a).
\end{equation}
The EoS parameter given by the approximation (\ref{cpl}) and also
what was obtained from our numerical solutions  are shown in fig.
(\ref{fig3})

\begin{figure}[H]
\centering\epsfig{file=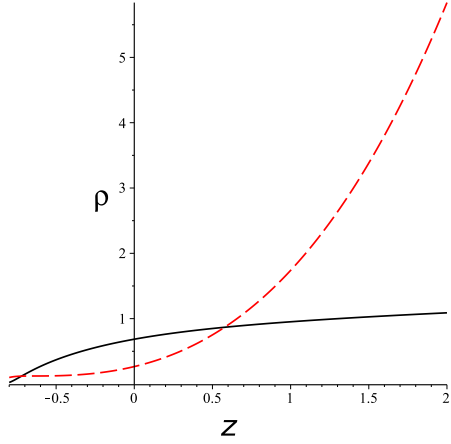,width=6cm,angle=0} \caption{
$\tilde{\rho_{cd}}$ (red dashed line) and $\tilde{\rho_{\varphi}}$
(black solid line) in terms of $z$ for $\{
\tilde{W}=1000,n=0.352,\tilde{\Gamma}=10000,
\Omega_{\varphi}(0)=0.685, \Omega_{cd}(0)=0.266,
\Omega_b(0)=0.049\}.$ Due to the interaction term, dark energy decreases faster than figure (\ref{fig1}) and dark matter becomes again dominant in the future.} \label{fig4}
\end{figure}

\begin{figure}[H]
\centering\epsfig{file=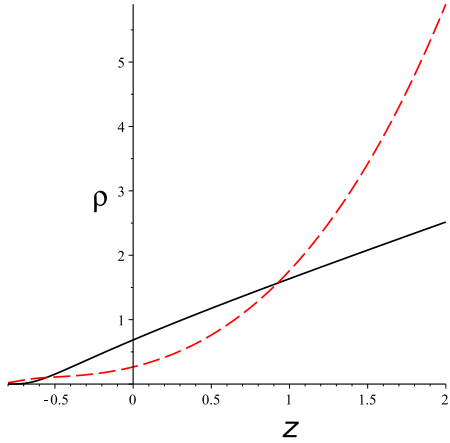,width=6cm,angle=0} \caption{
$\tilde{\rho_{cd}}$ (red dashed line ) and $\tilde{\rho_{\varphi}}$
(black solid line) in terms of $z$ and for $\{\tilde{\Gamma}=3, n=2,
\tilde{W}=1, \Omega_{\varphi}(0)=0.685, \Omega_{cd}(0)=0.266,
\Omega_b(0)=0.049 \}.$  the Presence of interaction term expedites the redomination of dark matter with respect to the figure (\ref{fig2}).} \label{fig5}
\end{figure}

In the presence of the interaction (\ref{26}), we plot the
solutions of the system in figs.(\ref{fig4}) and (\ref{fig5}).

We have still used the same initial conditions at $a=1$ as before.
By Comparing these figures with the previous ones, we find that in
the presence of $\Gamma$, the dark energy domination occurs at
higher redshift $z$. This is related to the fact that in the
presence of the energy exchange, the dark matter decreases with a
smaller slope while the dark energy decreases with a greater
slope.

A feature of our model,
which is shown clearly in the figures (\ref{fig4}) and
(\ref{fig5}), is that the matter may become dominant in
the future and the accelerated expansion ceases. We expect that the dark matter is diluted
faster than dark energy which has a more negative EoS parameter, so at first glance, it seems that the cease of acceleration or the dominance of dark matter at late time
is related to the positive decay rate of $\varphi$ to dark matter, but this is not the whole story: $H$ is a decreasing function of time, and the EoS parameter of $\varphi$, evolves from $w_\varphi=-{n+6\over 3n+6}$ at early time where $H^2\gg 1$, to $w_\varphi={n-2\over n+2}$ corresponding
to $WH^2\ll 1$ at late time.
So the EoS of $\varphi$  increases and this component may even behave as dark matter at late time (for n=2). In the era where
the Universe is nearly filled with dark energy, pressureless dark
matter and baryonic matter, the Universe EoS is be given by
\begin{eqnarray}
w&=&{P_\varphi\over \rho_\varphi+\rho_{cd}+\rho_b}\nonumber \\
&=&\Omega_\varphi w_\varphi.
\end{eqnarray}

\begin{figure}[H]
\centering\epsfig{file=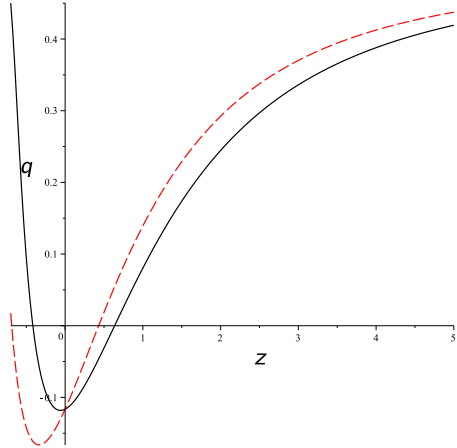,width=6cm,angle=0} \caption{ Deceleration
parameter of the Universe in the presence (black solid line) and absence
(red dashed line) of interaction with the same parameters and initial
conditions as figs.(\ref{fig2}) and (\ref{fig5}). Even in the absence of interaction, the Universe enters a deceleration phase in the future.
The presence of interaction expedites this procedure. } \label{fig7}
\end{figure}

$w>(<){-1\over 3}$ specifies deceleration (acceleration).  This can also be expressed in
terms of the deceleration parameter $q={1\over 2}(1+3w)$: for $q< (>)0$ the expansion is accelerating (decelerating).
In fig.(\ref{fig7}) the deceleration parameter is depicted in terms of the redshift. Employed parameters  are  $\{\tilde{\Gamma}=7, n=2,
\tilde{W}=1, \Omega_{\varphi}(0)=0.685, \Omega_{cd}(0)=0.266,
\Omega_b(0)=0.049 \}.$
This figure shows that even for $\Gamma=0$ the acceleration ceases in the future and the presence of interaction
term only expedites this procedure.

Note that in the aforementioned  examples, if one takes
$\Lambda\gg  10^{-5}eV$, where $\lambda=\Lambda^{4-n}$, the
condition required for rapid oscillation, i.e. (\ref{9}), is
satisfied. To obtain this, one can evaluate numerically the right
hand side of (\ref{9}) by taking $H_0=67.3 km s^{-1} Mpc^{-1}$
\cite{Planck}.
\section{Summary}
The non-minimal derivative coupling  model proposed in
\cite{sush,nonminimal}, was employed to describe the recent acceleration of
the Universe. In this context, instead of the well known slow roll
scalar field model, we considered a rapid oscillating scalar field with a power
law potential. After some preliminaries and generalizing the results of \cite{sad1} to the quintessence case,
we obtained required condition for such an oscillation and depending on the strength of
the coupling we divided our study into two categories. In the high
friction regime, we obtained a constant EoS parameter for dark
energy which was different from what was obtained in the minimal
case permitting the acceleration for a wider range of potentials.
In the intermediate regime, a dynamical EoS parameter was derived allowing
acceleration to deceleration phase transition in the future without invoking
interaction terms. This transition was discussed by obtaining a series solution for the system.
We also took into account a possible energy exchange
between dark components (inspired by inflaton decay during its
coherent rapid oscillation), alleviating the coincidence problem.
In the high friction regime, and the minimal model, where the EoS parameter of dark energy is a constant
the interaction term is responsible for the possible future acceleration to deceleration phase transition.
We showed also that when the dark energy decays to dark matter,
there exists a late time stable attractor for the system.
A short discussion was given on the instabilities of interacting dark energy model in the perturbed FRW Universe.
This discussion was restricted to high friction regime and the minimal model where the dark energy has a constant EoS parameter. An outlook is to investigate this problem in intermediate regime where the dark energy has a variable EoS parameter and also considering more general interactions terms proposed in the literature.

At the end, using numerical methods,  the evolution of dark sectors were depicted
and discussed in details.

\end{document}